\documentclass[12pt,twoside]{article}

\date{July 1, 2013} 

\begin{document} 

\centerline{\bf Adv. Studies Theor. Phys., Vol. x, 2013, no. xx, xxx - xxx} 

\centerline{\bf HIKARI Ltd, \ www.m-hikari.com}

\centerline{} 

\centerline{} 

\centerline {\Large{\bf Generalized Dirac and Klein-Gordon 
equations}} 

\centerline{} 

\centerline {\Large{\bf for spinor wavefunctions}} 

\centerline{} 

\centerline{\bf {R. Huegele, Z.E. Musielak and J.L. Fry}} 

\centerline{} 

\centerline{Department of Physics} 

\centerline{The University of Texas at Arlington} 

\centerline{Arlington, TX 76019, USA} 

\centerline{} 

\begin{abstract}
A novel method is developed to derive the original Dirac equation 
and demonstrate that it is the only Poincar\'e invariant dynamical 
equation for 4-component spinor wavefunctions.  New Poincar\'e 
invariant generalized Dirac and Klein-Gordon equations are also 
derived. In the non-relativistic limit the generalized Dirac 
equation gives the generalized L\'evy-Leblond equation and the 
generalized Pauli-Schr\"odinger equation.  The main difference 
between the original and generalized Dirac equations is that the 
former and latter are obtained with zero and non-zero phase 
functions, respectively.  Otherwise, both equations describe 
free elementary particles with spin 1/2, which have all other 
physical properties the same except their masses.  The fact that 
the generalized Dirac equation describes elementary particles with 
larger masses is used to suggest that non-zero phase functions may 
account for the existence of three families of elementary particles 
in the Standard Model.  This suggestion significantly differs from 
those previously made to account for the three families of particle 
physics. 
\end{abstract} 


\section{Introduction}

In the Special Theory of Relativity (STR), the background space-time is flat and 
endowed with the Minkowski metric.  All transformations of coordinates that 
leave the metric invariant form a representation of the inhomogeneous Lorentz 
group, which is also known as the Poincar\'e group [1,2].  Wigner [1] classified 
all irreducible representations (irreps) of the Poincar\'e group and used them 
to establish classes of elementary particles that exist in this space-time [2].  
A dynamical equation that is invariant with respect to the transformations of 
coordinates is a Poincar\'e invariant equation [3]. 

The Klein-Gordon (KG) equation [4,5], one of the fundamental (Poincar\'e invariant) 
equations of quantum field theories (QFT), describes bosons with spin zero that 
are represented by scalar wave functions.  Other fundamental equations of QFT 
are the Dirac [6] and Proca [7] equations, which describe fermions with spin 
1/2 and bosons with spin 1, respectively.  The wavefunctions that represent 
fermions are spinors while vector wavefunctions represent bosons.  To obtain 
the KG equation, the STR energy-momentum relationship is typically used and 
differential operators are substituted for the energy and momentum [8].  On 
the other hand, the Dirac equation is usually derived by Dirac's method [8] 
or by using the transformation properties of spinors under the Lorentz group 
[9].  A similar procedure is also used to obtain the Proca equation.  All these 
equations can also be formally obtained from an appropriate Poincar\'e invariant
Lagrangian density [10].  Another method was introduced by Bargmann and Wigner 
[11], who used the irreps of the Poincar\'e group to obtain the fundamental 
equations of QFT.  A complete description of the group theoretical derivation 
of the Dirac equation is presented by Thaller [12].  A different approach was 
developed by Fry, Musielak \& Chang [13], who formally derived the KG equation 
for free spin-zero particles using the principle of relativity and the principle 
of analyticity.

In Paper I of this series [14], we derived the L\'evy-Leblond equation [15,16] 
for a four-component spinor wavefunction as well as the corresponding Schr\"odinger 
equation, and proved that they were the only Galilean invariant four-component 
spinor equations with the Schr\"odinger phase function.  The relationship between 
these equations and the Pauli-Schr\"odinger equation [17] was discussed.  Extensive 
studies of L\'evy-Leblond and Pauli-Schr\"odinger equations were performed by 
Fushchich and Nikitin [3], who derived the equations and investigated their 
symmetries.  Moreover, a general method of constructing Galilean invariant 
theories was developed by de Montigny et al. [18] and Niederle and Nikitin [19].  
A different method based on the principle of relativity and the principle of 
analyticity was developed by Musielak and Fry [20] and by Fry and Musielak [21].   

Using different phase functions than the Schr\"odinger phase function considered
in Paper I, we derived new Galilean invariant dynamical equations, which were 
called the generalized L\'evy-Leblond and generalized Schr\"odinger equations 
[22]; these equations reduce to the standard L\'evy-Leblond and Schr\"odinger 
equations when the Schr\"odinger phase function was used.  We demonstrated that 
the standard and generalized equations described the same elementary particle
with spin $1/2$ but with different masses; the mass was larger for the generalized
equation.  Despite that all results were obtained in the non-relativistic regime
[22], we suggested that the difference in mass resulting from using different 
phase functions may account for the existence of three families of elementary 
particles in the Standard Model of particle physics [23].  

One of the main goals of this paper is to verify the above suggestion in the 
relativistic regime.  In order to achieve this goal, we first develop a novel 
method of deriving the Dirac equation, and then use this method to search for 
new fundamental (Poincar\'e invariant) dynamical equations.  The result of 
this search is the generalized Dirac equation for four-component spinor 
wavefunctions.  To check the validity of this generalized Dirac equation, 
we evaluated it in the non-relativistic limit and showed that the generalized 
L\'evy-Leblond equation [22] and the generalized Pauli-Schr\"odinger equation 
are obtained.  Moreover, we also derive a second order equation, the generalized 
Klein-Gordon (KG) equation.  The new generalized Dirac and Klein-Gorodon equations
are obtained by using non-zero phase factors, and these equations reduce to the 
standard Dirac and Klein-Gordon equations once the phase function is set to zero. 

The outline of this paper is as follows: governing equations are given in 
section 2; a novel derivation of the Dirac equation is presented in section 3;
the generalized Dirac and Klein-Gordon equations are derived in sections 4 
and 5, respectively; the generalized L\'evy-Leblond and Pauli-Schr\"odinger 
equations are obtained in the non-relativistic limits of the generalized Dirac
equations in section 6; the generalized Dirac equation and the existence of
three familes of elemementary particles in the Standard Model are discussed
in section 7; and our conclusions are presented in section 8.   

\section{Governing equations}

The Minkowski metric can be written as $ds^2\ =\ dt^2 - dx^2 - dy^2 - 
dz^2$, where $x$, $y$ and $z$ are the spatial coordinates and $t$ is the 
time coordinate given in natural units where the speed of light is $c = 1$.  
The group of this metric is the Poincar\'e group $P$, whose structure is 
given by the following semi-direct product: $P\ =\ H_p \otimes_s T (3+1)$, 
where $T(3+1)$ is an invariant subgroup of space-time translations and $H_p$ 
is a non-invariant subgroup consisting of the remaining transformations 
and the identity transformation.  In this paper, we consider the proper 
orthochronous group $P^{\uparrow}_{+}$ that is a subgroup of $P$.

The semi-direct product structure of the Poincar\'e group guarantees that 
a given wavefunction $\psi (x, t)$ must transform like an irrep of the 
invariant subgroup $T(3+1)$.  Thus a necessary condition that a wave 
described by $\psi$ represents an elementary particle [13] is 
\begin{equation}
i \partial_{\mu} \psi (x,t)\ =\ k_{\mu} \psi (x,t)\ ,
\label{eq1}
\end{equation}

\noindent
where $\partial_{\mu} = \partial / \partial x^{\mu} = ( \partial / c \partial t, 
\nabla )$ and $k_{\mu} = ( {\omega / c}, - k_x, - k_y, - k_z )$.  This equation 
must be satisfied for each component of a multi-component wavefunction.

Since in this paper, we consider elementary particles with spin $1/2$, their 
wavefunctions are spinors.  With the assumption that $\psi$ is a spinor, Eq.
(\ref{eq1}) must be modified to account for different components of the spinor
wavefunction.  Eq. (\ref{eq1}) is not invariant, therefore, we seek an invariant
version that remains linear in its derivatives while invariant to all Poincar\'e
transformations.  This can be achieved by multiplying both sides of Eq. (\ref{eq1})
by an arbitrary, constant matrix $A^{\mu}$, which in general has dimension $n \times 
n$ to be consistent with the $n$-component wavefunction (see below).  The result is 

\begin{equation}
i A^{\mu} \partial_{\mu} \psi (x,t)\ =\ A^{\mu} k_{\mu} \psi (x,t)\ .
\label{eq2}
\end{equation}

Defining $B^{\mu} \equiv i A^{\mu}$ and $B_c \equiv A^{\mu} k_{\mu}$, we 
obtain the first order differential equation of the following form
\begin{equation}
\left [ B^\mu \partial_\mu + B_c \right ] \psi(x,t) = 0\ . 
\label{eq3}
\end{equation}

\noindent
We shall now determine conditions that the matrices $B^{\mu}$ and $B_c$ 
must satisfy in order for the above equation to be Poincar\'e invariant.  
The requirement is that the obtained dynamical equations for $\psi$ 
remain the same in all inertial frames of reference that exist in 
Minkowski space-time, as required by the principle of relativity. 

It must be noted that we seek an equation whose solutions have eigenvalues
corresponding to spin $1/2$ for the second Casimir operator of the Poincar\'e
group [2].  We also seek solutions that correspond to unitary irreps with
spin $1/2$ that are given by the covering group of $P^{\uparrow}_{+}$, which 
is a group of $4 \times 4$ matrices [2,12].  Thus, a wavefunction for a unitary 
spin $1/2$ irrep must have at least $4$ components, so $n = 4$.  

\section{Novel derivation of Dirac equation}

\subsection{Conditions for invariance of the first order equation}

A space-time point in the coordinates of one observer $x^\nu$ is related 
to the same point in the coordinates of another observer $x'^\nu$ by a 
linear transformation 
\begin{equation}
x'^{\nu}=\Lambda^\nu_\mu x^\mu + b^\nu\ .
\label{eq4}
\end{equation}

\noindent
Under the linear transformation a differential operator transforms like
\begin{equation}
\partial'_\mu = \Lambda^\rho_\mu \partial_\rho\ ,
\label{eq5}
\end{equation}

\noindent 
and the wavefunction transforms like
\begin{equation}
T\psi(x,t) =\psi'(x',t')\ .
\label{eq6}
\end{equation} 

\noindent
Moreover, since we require $\vert \psi \vert^2 = \vert \psi' \vert^2$, 
the wavefunctions can be related by a phase factor 
\begin{equation}
 e^{i\phi(x,t)}\psi(x,t) =\psi'(x',t')\ .
\label{eq7}
\end{equation} 

The principle of relativity requires that if a dynamical equation such as 
Eq. (\ref{eq3}) is true for one observer, then there exists another equation 
of the same form for all other observers who have a coordinate system that 
is translated, rotated, and boosted relative to the first observer.  An 
arbitrary second observer would have an equation that would feature objects 
such as $B'^\mu$, $B_c'$, and $\partial'_\mu$, which may have been changed 
by the coordinate transformation such as 
\begin{equation}
\left[B'^\mu \partial'_\mu + B'_c\right]\psi'(x',t') = 0\ .
\label{eq8}
\end{equation} 

Substituting the transformation rules of the differential operator (see Eq. 
\ref{eq5}) and the wavefunction (see Eq. \ref{eq6}) into Eq. (\ref{eq7}) 
produces a first order equation that is now written in terms of the original 
coordinate system $x^\mu$
\begin{equation}
\left[ B'^\mu \Lambda^\rho_\mu \partial_\rho  +  B'_c \right]e^{i 
\phi(x,t)} \psi (x, t) = 0\ .
\label{eq9}
\end{equation}

\noindent
The phase factor can be commuted through the differential operator and the 
result is 
\begin{equation}
\Lambda^\rho_\mu \partial_\rho  e^{i \phi(x,t)} = e^{i \phi(x,t)} 
\Lambda^\rho_\mu \left( \partial_\rho + i \partial_\rho \phi \right)\ .
\label{eq10}
\end{equation}

\noindent
Commuting the phase factor through to the left side of the transformed first 
order equation and dividing it out, we obtain
\begin{equation}
\left[ B'^\mu  \Lambda^\rho_\mu \left( \partial_\rho + i \partial_\rho \phi 
\right)  +  B'_c \right] \psi (x, t) = 0\ .
\label{eq11}
\end{equation}

For the first order equation to be invariant, it is required that the transformed 
first order equation is the same as the original first order equation.  Equating 
terms of like differential powers then generates a set of conditions on the 
matrices $B^\mu$ and $B_c$ that must be met for the dynamical equation to be 
invariant.  The conditions are:  
\begin{equation}
B^\mu=B'^\beta \Lambda^\mu_\beta\ ,
\label{eq12}
\end{equation}
and
\begin{equation}
B_c=B'_c+iB'^\beta \Lambda^\mu_\beta \partial_\mu \phi\ .
\label{eq13}
\end{equation}

\noindent
The equation is invariant for phase functions of the form
\begin{equation}
\phi(x,t)= \zeta_\mu x^\mu + \zeta_c\ ,
\label{eq14}
\end{equation}

\noindent
where $\zeta_\mu$, $\zeta_c$ are scalar functions of the transformation 
parameters $a^\mu$, $v^i$, and $\theta^i$ and potentially any number of 
other parameters that have yet to be introduced.  Using the replacement 
$\partial_\mu \phi = \zeta_\mu$, the condition for invariance (see Eq. 
\ref{eq13}) becomes
\begin{equation}
B_c=B'_c+iB'^\beta \Lambda^\mu_\beta \zeta_\mu\ .
\label{eq15}
\end{equation}
%

%%%%%%%%%%%%%%%%%%%%%%%%%%%%%%%%%%%%%%%
\subsection{Matrices that satisfy the conditions for invariance}

The matrices $B^\mu$ and $B_c$ must satisfy the conditions (see Eqs \ref{eq12} 
and \ref{eq15}) in order to form a first order differential equation that is 
invariant under the linear transformation  $\hat {\Lambda}$.  From this point 
on the linear transformation will be a Poincare or inhomogeneous Lorentz 
transformation.  This transformation can be separated into rotations and 
boosts to ease calculations.  The rotations matrices for 4-component spinors 
are:
\begin{equation}
R_{\theta_j}= \cos {\theta \over 2} + \epsilon_{jkl} \gamma_k \gamma_l 
\sin {\theta \over 2}\ ,
\label{eq16}
\end{equation}

\noindent
where $\gamma_j$ and $\gamma_0$ form a basis for the $4 \times 4$ spinor 
matrices and $\theta_j$ is the rotation about the $j$-axis and in the 
$k-l$-plane.  The boost matrices for 4-component spinors are:
\begin{equation}
S_{v_j}=\cosh {{\eta \over 2}}+i \gamma_j \gamma_0 \sinh {{\eta \over 2}}\ ,
\label{eq17}
\end{equation}

\noindent
where $\eta$ is the boost angle.  The boost angle is related to the velocity by 
$\tanh {\eta} = \beta=v/c$ where $c=1$ is the speed of light in natural units. 
 
The Dirac representation is typically chosen whenever explicit representation 
of the gamma matrices $\gamma^\mu$ is required 

\begin{equation}
\gamma^j = \left( \begin{array}{cc}
0 & i\sigma^j \\ 
-i\sigma^j & 0 \end{array}
\right)\ ,
\ \  \ \ 
\gamma^0 =  \left( \begin{array}{cc}
I & 0 \\ 
0 & -I \end{array}
\right)\ ,
\label{eq18}
\end{equation}

\noindent
where $\sigma^j$ are the standard Pauli matrices
\begin{equation}
\sigma^x=\left( \begin{array}{cc}
0 & 1 \\
1 & 0 \end{array}
\right)\ ,
\ \ \ \ 
\sigma^y=\left( \begin{array}{cc}
0 & -i \\
i & 0 \end{array}
\right)\ ,
\ \ \ \ 
\sigma^z=\left( \begin{array}{cc}
1 & 0 \\
0 & -1 \end{array}
\right)\ ,
\label{eq19}
\end{equation}

\noindent
In this representation the Minkowski metric $\eta^{\mu \nu}$ has signature 
$(+ - - -)$.  The covariant gamma matrices are related to the contravariant 
form by $\gamma_\mu = \eta_{\mu \nu} \gamma^\nu = \{\gamma^0, -\gamma^j\}$.  
The gamma matrices satisfy the Clifford algebra defined by the anti-commutators
\begin{equation}
\left\{ \gamma^\mu , \gamma^\nu \right\} = 2 \delta^{\mu \nu}I\ .
\label{eq20}
\end{equation}

\subsection{Invariant equations for the phase function $\phi=0$}

Before dealing with the more general case we consider the consequences of 
assuming the phase function $\phi(x,t)=0$.  This case requires $\zeta_\mu = 
\zeta_c = 0$ and simplifies the condition (see Eq. \ref{eq15}) to $B_c=B'_c$.
Now, we apply rotations, which constrains the matrices to the form
\begin{equation}B^t=\left( \begin{array}{cc}
pI & qI \\ 
sI & tI \end{array}
\right)\,  \  \   B^{j}=\left( \begin{array}{cc}
e{\sigma }^j & f{\sigma }^j \\ 
g{\sigma }^j & h{\sigma }^j \end{array}
\right)\ ,
\label{eq21} 
\end{equation} 

\noindent 
where $e, f, g, h, p, q, s$, and $t$ are unknown scalar constants.  

We re-write condition (see Eq. \ref{eq13}) with the boost transformation $S$,
and obtain
\begin{equation}
\Lambda^\beta_\mu B^\mu =B'^\beta = S B^\beta S^{-1}\ ,
\label{eq22}
\end{equation}

\noindent
Applying a Lorentz boost along the $z$-direction in the above condition,  
the RHS becomes
\[
S_z B^\beta S_z^{-1}=\left( \cosh {{\eta \over 2}}-i \gamma_z \gamma_0 
\sinh {{\eta \over 2}} \right)B^\beta \left( \cosh {{\eta \over 2}}+ i 
\gamma_z \gamma_0 \sinh {{\eta \over 2}}\right)
\]
\[
= B^\beta \cosh ^2 {\eta \over 2}+\gamma_z \gamma_0 B^\beta \gamma_z 
\gamma_0 \sinh^2 {\eta \over 2} 
\]
\[
\ \ \ \ +i \left( B^\beta \gamma_z \gamma_0- \gamma_z \gamma_0 B^\beta 
\right) \sinh {\eta \over 2} \cosh {\eta \over 2}
\]
\[
={1 \over 2}\left(B^\beta + \gamma_z \gamma_0 B^\beta \gamma_z \gamma_0 
\right) \cosh \eta + {1 \over 2}\left( -B^\beta + \gamma_z \gamma_0 B^\beta 
\gamma_z \gamma_0 \right)
\]
\begin{equation}
\ \ \ +{i \over 2} \left( B^\beta \gamma_z \gamma_0 - \gamma_z \gamma_0 
B^\beta \right) \sinh \eta\ .
\label{eq23}
\end{equation}

While the LHS of Eq. (\ref{eq22}) is
\begin{equation}
\Lambda^t_\mu B^\mu = B^t \cosh \eta - i B^z \sinh \eta\ ,
\label{eq24}
\end{equation}
\noindent
for $\beta = t$ and a boost in the $z$-direction.  Alternatively the LHS is
\begin{equation}
\Lambda^z_\mu B^\mu =B^z \cosh \eta + i B^t \sinh \eta\ ,
\label{eq25}
\end{equation}

\noindent
for $\beta = z$ while $\Lambda^x_\mu B^\mu =B^x$ and $\Lambda^y_\mu B^\mu = B^y$.
Here we used the following identities
\begin{equation}
\sinh^2 {\eta \over 2} ={1 \over 2} \cosh \eta + {1 \over 2}\ ,
\label{eq26a}
\end{equation}
and
\begin{equation}
\sinh {\eta \over 2} \cosh {\eta \over 2} = {1 \over 2} \sinh \eta\ ,
\label{eq26b}
\end{equation}
and
\begin{equation}
\cosh^2{\eta \over 2} = {1 \over 2} \cosh \eta - {1 \over 2}\ .
\label{eq26c}
\end{equation}

The RHS of Eq. (\ref{eq22}) is equal the LHS when the following 
constraints:
$\left\{ B^t, \gamma_z \gamma_0 \right\}=0$, 
$\left\{ B^z, \gamma_z \gamma_0 \right\}=0$,
$\left[ B^x, \gamma_z \gamma_0 \right]=0$,
and $\left[ B^y, \gamma_z \gamma_0 \right]=0$
are obeyed.

\noindent
The above procedure may be repeated for boosts in the $x$ and $y$ directions 
to produce a set of constraints similar to those given above.  Taken together 
the constraints are equivalent to requiring the matrices $B^\mu$ obey the 
Clifford algebra: $\left\{B^\mu, B^\nu \right\} = 2\delta^{\mu \nu}I$.

We now return to the condition on $B_c = B'_c$.  Application of a Lorentz boost 
gives
\[
B_c=S_z B_c S_z^{-1}
\]
\[
=\left( \cosh {{\eta \over 2}}-i \gamma_z \gamma_0 \sinh {{\eta \over 2}} \right)
B_c \left( \cosh {{\eta \over 2}}+i \gamma_z \gamma_0 \sinh {{\eta \over 2}}\right)
\]
\[
= B_c \cosh ^2 {\eta \over 2}+\gamma_z \gamma_0 B_c \gamma_z \gamma_0 \sinh^2 
{\eta \over 2} 
\]
\begin{equation}
\ \ \ \ +i \left( B_c \gamma_z \gamma_0- \gamma_z \gamma_0 B_c \right) \sinh 
{\eta \over 2} \cosh {\eta \over 2}\ ,
\label{eq27}
\end{equation}

\noindent
which leads to $\left[ B_c, \gamma_z \gamma_0 \right] = 0$.  However, application 
of rotations gives
\[
B_c=R_zB_cR_z^{-1}=\left( \cos {{\theta \over 2}}- \gamma_x \gamma_y \sin 
{{\theta \over 2}} \right)B_c \left( \cos {{\theta \over 2}}+ \gamma_x 
\gamma_y \sin {{\theta \over 2}}\right)
\]
\begin{equation}
=B_c \cos^2 {\theta \over 2} - \gamma_x \gamma_y B_c \gamma_x \gamma_y 
\sin^2 {\theta \over 2} + \left(B_c \gamma_x \gamma_y - \gamma_x 
\gamma_y B_c \right) \sin {\theta \over 2} \cos {\theta \over 2}\ ,
\label{eq28}
\end{equation}

\noindent
and leads to the constraint $\left[ B_c, \gamma_x \gamma_y \right] = 0$.

This above reasoning can be repeated for $x$- and $y$-directions, to 
produce similar constraints.  Under these conditions the matrix $B_c$ is 
constrained to
\begin{equation}
B_c=a\left( \begin{array}{cc}
I & 0 \\ 
0 & I \end{array}
\right)\ ,
\label{29}
\end{equation}

\noindent
where $a$ remains a free parameter.

\subsection{Dirac equation}

Combining all the above results, we obtain
\begin{equation}
\left[
 \left( \begin{array}{cc}
0 & \sigma^j \\ 
-\sigma^j & 0 \end{array}
\right) \partial_j 
+
 \left( \begin{array}{cc}
I & 0 \\ 
0 & -I \end{array}
\right) \partial_t
+
a\left( \begin{array}{cc}
I & 0 \\ 
0 & I \end{array}
\right)
\right]\psi(x,t) = 0\ ,
\label{eq30a}
\end{equation}

\noindent
which is a Poincar\'e invariant equation for 4-component spinor state functions.  
This equation satisfies the principles of relativity and analyticity but we must 
also insure that the state functions transform like irreducible representations 
of the Poincar\'e group.  This means that the state functions must obey the 
eigenvalue equations $i\partial_t \psi = \omega \psi$ and $-i\partial_j \psi = 
k_j \psi$, with $j = 1$, $2$ and $3$ (see Eq. \ref{eq1}).  The requirement can 
be used to remove the free parameter $a$.  Multiplying Eq. (\ref{eq30a}) by $i$ 
and moving the time derivative to the LHS yields
\begin{equation}
i \left( \begin{array}{cc}
I & 0 \\ 
0 & -I \end{array}
\right) \partial_t \psi(x,t)
=-i\left[
 \left( \begin{array}{cc}
0 & \sigma^j \\ 
-\sigma^j & 0 \end{array}
\right) \partial_j 
+
a\left( \begin{array}{cc}
I & 0 \\ 
0 & I \end{array}
\right)
\right]\psi(x,t)\ .
\label{eq30b}
\end{equation}

\noindent
Clearly, the operator on the right hand side is the Hamiltonian and for $psi$ 
to transform as a unitary spin $1/2$ irrep it must be Hermitian.  The terms 
are all Hermitian under the condition that the constant $a$ is imaginary or 
that $ia$ is real.  Now we substitute the eigenvalues for their operators, 
and obtain
\begin{equation}
\omega \left( \begin{array}{cc}
I & 0 \\ 
0 & -I \end{array}
\right)\psi(x,t)
= k_j\left( \begin{array}{cc}
0 & \sigma^j \\ 
-\sigma^j & 0 \end{array}
\right)\psi(x,t)
-ia\left( \begin{array}{cc}
I & 0 \\ 
0 & I \end{array}
\right)
\psi(x,t)\ .
\label{eq30c}
\end{equation}

Substituting the bispinor
\begin{equation}
\psi (x,t)=\left( \begin{array}{c} \phi \\ \eta \end{array} \right)\ ,
\label{31}
\end{equation}

\noindent
for the wavefunction in Eq. (\ref{eq30c}), we obtain the following pair of 
linked equations:
\begin{equation}
\omega \phi = k_j \sigma_j \eta -ia\phi\ , 
\label{eq32a}
\end{equation}
and
\begin{equation}
-\omega \eta = -k_j \sigma_j \phi -ia\eta\ .
\label{eq32b}
\end{equation}

The four momentum $p_\mu$ contains the eigenvalues of energy and momentum
$p_\mu=\left( \omega, k_j \right)$.  There always exists some frame of 
reference where the momentum is zero ($k_j = 0$) and the energy equals the 
rest mass, so that $\omega=\omega_0$ where $\omega_0$ is called the invariant 
frequency.  When these values are substituted into the pair of equations the 
only remaining free parameter $a$ can be determined in terms of the invariant 
frequency $ia=\pm \omega_0$ and Eq. (\ref{eq30c}) can be written as
\begin{equation}
i \left( \begin{array}{cc}
I & 0 \\ 
0 & -I \end{array}
\right) \partial_t \psi(x,t)
=\left[
-i \left( \begin{array}{cc}
0 & \sigma^j \\ 
-\sigma^j & 0 \end{array}
\right) \partial_j 
\pm\omega_0
\left( \begin{array}{cc}
I & 0 \\ 
0 & I \end{array}
\right)\right]
\psi(x,t)\ . 
\label{eq30d}
\end{equation}

\noindent
By identifying the wave mass with the rest mass, the derived equation becomes
the Dirac equation [6].  Note that the above derivations were performed with 
the phase function $\phi=0$, consequently this is the only Poincare invariant 
first-order differential equation for $\phi=0$.  It is also important to point 
out that our derivation of the Dirac equation is new and significantly different 
than typical derivations shown in QFT textbooks [8,9]. 
%

%%%%%%%%%%%%%%%%%%%%%%%%%%%%%%%%%%%%%%%%%%%%%%%%%%%%%
\section{Generalized Dirac equation}

We now address the more general case when the phase function is other than 
$\phi=0$.  Since $\phi$ is not present in the condition for $B^\mu$ (see Eq. 
\ref{eq13}), there is no impact on the results obtained for $B^\mu$ in the 
previous section.  The matrix $B_c$ however is constrained by Eq. (\ref{eq15}), 
where the presence of $\phi$ introduces the function $\zeta_\mu(\theta_i, v_j)$.  
Now the problem is to determine the value of the functions $\zeta_\mu(\theta_i, 
v_j)$ in addition to the matrix $B_c$.  

First notice that the condition given by Eq. (\ref{eq15}) can be simplified by 
the substitution of condition given by Eq. (\ref{eq12}) and the use of the result $B^\mu=\gamma^\mu$.  This gives
\begin{equation}
B_c=B'_c+iB'^\beta \Lambda^\mu_\beta \zeta_\mu = B'_c+iB^\mu \zeta_\mu= 
B'_c+i\gamma^\mu \zeta_\mu\ .
\label{eq33}
\end{equation}

\noindent
By setting $\theta_j=0$ and $v_j=0$ it is clear that $\zeta_\mu(0,0)=0$.  
To find the rotational dependence we move the terms containing $B_c$ to 
the LHS and apply a z-rotation.  The result is 
\[
B_c-B'_c=\left(B_c+\gamma_1\gamma_2B_c\gamma_1\gamma_2\right)\sin^2{\theta_3 
\over 2} 
\]
\begin{equation}
- \left(B_c\gamma_1\gamma_2-\gamma_1\gamma_2B_c\right)\sin{\theta_3 \over 2}
\cos{\theta_3 \over 2}\ ,
\label{eq34}
\end{equation}

\noindent
where we used the following identities
\begin{equation}
\sin\theta_3 = 2 \sin {\theta_3 \over 2} \cos {\theta_3 \over 2}\ , 
\ \ \ \ \ \   \cos \theta_3 =  \cos^2 {\theta_3 \over 2}-\sin^2
{\theta_3 \over 2}\ ,
\label{eq35a}
\end{equation}
and
\begin{equation}
{1-\cos \theta_3 \over 2} = \sin^2 {\theta_3 \over 2}\ .
\label{eq35b}
\end{equation}

Any $4\times4$ complex matrix can be represented using the matrices $I$, 
$\gamma^\mu$, $\sigma^{\mu\nu}$, $\gamma^5\gamma^\mu$, and $\gamma^5$ as 
a basis where 
\begin{equation}
\sigma^{\mu\nu}={i \over 2}\left[\gamma^\mu,\gamma^\nu\right]\ ,
\label{eq36a}
\end{equation}
\begin{equation}
\gamma^5=i\gamma^0\gamma^1\gamma^2\gamma^3=
-i \left( \begin{array}{cc}
0 & I \\ 
I & 0 \end{array}
\right)
\label{eq36b}
\end{equation}
and
\begin{equation}
\gamma^5\gamma^j=
 \left( \begin{array}{cc}
-\sigma^j & 0 \\ 
0 & -\sigma^j \end{array}
\right)
\ \ \ \ \  \ \ 
\gamma^5\gamma^0=
i\left( \begin{array}{cc}
0 & I \\ 
-I & 0 \end{array}
\right)\ ,
\label{eq36c}
\end{equation}

\noindent
are all given in the Dirac basis.  If a matrix is required to be Hermitian, 
then the basis can be restricted to $I$, $\gamma^\mu$, $\sigma^{\mu\nu}$, 
and $\gamma^5\gamma^\mu$ excluding the anti-Hermitian $\gamma^5$.  Since 
the matrix $B_c$ must be Hermitian, it can be written in a basis composed 
of the Hermitian matrices
\begin{equation}
B_c = aI +b_{\mu\nu}\sigma^{\mu\nu} + c_\mu \gamma^\mu + d_\mu
\gamma^5\gamma^\mu\ ,
\label{eq37a}
\end{equation}
where
\begin{equation}
 c_\mu \gamma^\mu = c_0 \gamma_0 - c_1 \gamma_1 -c_2 \gamma_2 -c_3 \gamma_3\ ,
\label{eq37b}
\end{equation}

\noindent
with $a$, $b_i$, $c_\mu$, and $d_\mu$ are undetermined constants.  Using this 
basis we can calculate $B_c+\gamma_x \gamma_y B_c \gamma_x \gamma_y$ and $B_c 
\gamma_x \gamma_y - \gamma_x \gamma_y B_c$.  We do this for the $aI$ term first
and obtain
\begin{equation}
aI+\gamma_1\gamma_2aI\gamma_1\gamma_2=0\ ,
\label{eq38a}
\end{equation}
and
\begin{equation}
aI\gamma_1\gamma_2-\gamma_1\gamma_2aI=0\ .
\label{eq38b}
\end{equation}

Since these quantities vanish, $a$ is allowed to remain a free parameter; 
this guarantees that the Dirac equation is obtained as a special case of 
$\phi = 0$.  For the $b_{\mu\nu}\sigma^{\mu\nu}$ term, we find
\[
b_{\mu\nu}\sigma^{\mu\nu}+\gamma_1\gamma_2b_{\mu\nu}\sigma^{\mu\nu}
\gamma_1\gamma_2
\]
\begin{equation}
=i\left(-2b_{10}\gamma_0\gamma_1-b_{20}\gamma_0\gamma_2-b_{12}+
b_{12}\gamma_1\gamma_2+2b_{23}\gamma_2\gamma_3-2b_{31}\gamma_1\gamma_3 
\right)\ ,
\label{39a}
\end{equation}
and
\begin{equation}
b_{\mu\nu}\sigma^{\mu\nu}\gamma_1\gamma_2-\gamma_1\gamma_2b_{\mu\nu}
\sigma^{\mu\nu}=-2i\left(b_{10}\gamma_0\gamma_2-b_{23}\gamma_1\gamma_3
-b_{31}\gamma_2\gamma_3 \right)\ .
\label{39b}
\end{equation}

\noindent
It is apparent that all but the $b_{30}$ term must equal zero.  For the 
$c_\mu \gamma^\mu$ term, we have
\begin{equation}
 c_\mu \gamma^\mu+\gamma_x \gamma_y c_\mu \gamma^\mu \gamma_x \gamma_y = -2c_1\gamma_1-2c_2\gamma_2\ ,
\label{eq40a}
\end{equation}
and
\begin{equation}
c_\mu \gamma^\mu \gamma_x \gamma_y - \gamma_x \gamma_y  c_\mu \gamma^\mu = 
-2 c_1 \gamma_2 + 2 c_2 \gamma_1\ .
\label{eq40b}
\end{equation}

\noindent
The above relationships will survive since they are coefficients of individual 
$\gamma$-matrices, which appear in $\gamma^mu\zeta_mu$.  The $d^\mu \gamma_5
\gamma_\mu$ term produces
\begin{equation}
d^\mu\gamma_5\gamma_\mu+\gamma_1\gamma_2d^\mu\gamma_5\gamma_\mu
\gamma_1\gamma_2=-2id_1\gamma_0\gamma_2\gamma_3+2id_2\gamma_0
\gamma_1\gamma_3\ ,
\label{eq41a}
\end{equation}
and
\begin{equation}
d^\mu\gamma_5\gamma_\mu\gamma_1\gamma_2-\gamma_1\gamma_2d^\mu\gamma_5
\gamma_\mu=2id_1\gamma_0\gamma_1\gamma_3+2id_2\gamma_0\gamma_3\ .
\label{41b}
\end{equation}

\noindent
However, these terms must vanish to satisfy invariance of the first order 
equation.  Thus, we conclude that $d_1=0$ and $d_2=0$.

Putting the above results together, we have
\begin{equation}
\gamma_x \gamma_y B_c \gamma_x \gamma_y = -c_0 \gamma_0 -c_1 \gamma_1 - 
c_2 \gamma_2 + c_3 \gamma_3\ ,
\label{eq42a}
\end{equation}
and
\begin{equation}
B_c \gamma_x \gamma_y - \gamma_x \gamma_y B_c = -2 c_1 \gamma_2 + 
2 c_2 \gamma_1\ .
\label{eq42b}
\end{equation}

The remaining terms must vanish and thus provide a means of eliminating 
some of the free parameters of Eq. (\ref{eq37a}).  Inserting these results 
into the LHS of Eq. (\ref{eq34}), we have
\[
i\gamma^\mu\zeta_\mu=B_c-B'_c=\left(-2c_1\gamma_1-2c_2\gamma_2 \right)
\sin^2{\theta_3 \over 2}
\]
\[
-\left( -2 c_1 \gamma_2 + 2 c_2 \gamma_1\right)\sin{\theta_3 \over 2}
\cos{\theta_3 \over 2}
\]
\begin{equation}
=\left(-c_1\gamma_1-c_2\gamma_2 \right)(1-\cos\theta_3)+(c_1\gamma_2-
c_2\gamma_1)\sin\theta_3\ .
\label{eq43}
\end{equation}

\noindent
The $\gamma$-matrices are linearly independent so the coefficients of 
each matrix must vanish independently.  For a z-rotation we conclude 
that $\zeta_3(\theta_3)=0$ and $\zeta_0(\theta_3)=0$ and
\begin{equation}
\zeta_1 = i c_1 \left(1-\cos\theta_3\right)+ic_2 \sin\theta_3\ ,
\label{eq44a}
\end{equation}
and
\begin{equation}
\zeta_2 = i c_2 \left(1-\cos\theta_3\right)-ic_1\sin\theta_3\ .
\label{eq44b}
\end{equation}

\noindent
Similar expressions can be produced for the dependence of $\zeta_1$ 
and $\zeta_2$ on the angles $\theta_1$ and $\theta_2$.  All the free 
parameters of $B_c$ in Eq. (\ref{eq37a}) will vanish except for the 
$c_\mu$ terms and $a$.

The phase function $\phi$ will also depend on the boost parameters $v_j$.  
To find the functional dependence of $\phi$ on $v_j$, we apply boosts 
in Eq. (\ref{eq15}) and attempt to solve for $\phi$.  Equation (\ref{eq15}) 
is simplified by substituting in Eq. (\ref{eq12}) to get
\begin{equation}
B_c=B'_c+iB'^\beta \Lambda^\mu_\beta \zeta_\mu = B'_c+iB^\beta 
\zeta_\beta\ .
\label{45}
\end{equation}

\noindent
Applying a Lorentz boost along the $z$-direction produces
\[
B_c - S_z B_c S_z^{-1}=B_c-\left( \cosh {{\eta \over 2}}-i \gamma_z 
\gamma_0 \sinh {{\eta \over 2}} \right)
\]
\[
\times B_c \left( \cosh {{\eta \over 2}}+i \gamma_z \gamma_0 \sinh 
{{\eta \over 2}}\right)
\]
\[
= B_c-B_c \cosh ^2 {\eta \over 2}-\gamma_z \gamma_0 B_c \gamma_z 
\gamma_0 \sinh^2 {\eta \over 2} 
\]
\begin{equation}
\ \ \ \ -i \left( B_c \gamma_z \gamma_0- \gamma_z \gamma_0 B_c \right) 
\sinh {\eta \over 2} \cosh {\eta \over 2}\ .
\label{eq46}
\end{equation}

\noindent
With the remaining free parameters and $B_c=aI+c^\nu\gamma_\nu$, we 
calculate
\begin{equation}
B_c\gamma_3 \gamma_0 -\gamma_3 \gamma_0 B_c = -2(c_0\gamma_3+c_3
\gamma_0)\ ,
\label{47a}
\end{equation}
and
\begin{equation}
\gamma_3\gamma_0B_c\gamma_3\gamma_0=c_0\gamma_0+c_1\gamma+c_2
\gamma_2-c_3\gamma_3\ .
\label{47b}
\end{equation}

\noindent
Then Eq. (\ref{eq46}) can be written as
\[
i\gamma^\mu \zeta_mu=i\left[\gamma_0\zeta_0-\gamma_1\zeta_1-\gamma_2
\zeta_2-\gamma_3\zeta_3 \right]\ ,
\]
\[
=\gamma_0\left[ -2c_0\sinh^2 {\eta \over 2}+2ic_3\sinh{\eta\over2}
\cosh{\eta\over2}\right]
\]
\begin{equation}
+\gamma_3\left[2c_3\sinh^2{\eta\over2}+2ic_0\sinh{\eta\over2}
\cosh{\eta\over2}\right]\ ,
\label{eq48}
\end{equation}

\noindent
and the z-boost dependence of $\zeta$ is $\zeta_1=0$ and $\zeta_2=0$, 
which gives
\begin{equation}
\zeta_0= 2ic_0\sinh^2{\eta\over2}+2c_3\sinh{\eta\over2}
\cosh{\eta\over2}\ ,
\label{eq49a}
\end{equation}
and
\begin{equation}
\zeta_3=2ic_3 \sinh^2 {\eta\over2}-2c_0\sinh{\eta\over2}
\cosh{\eta\over2}\ .
\label{eq49b}
\end{equation}

\noindent
Equations (\ref{eq49a}) and (\ref{eq49b}) show how the phase function 
$\phi$ depends upon a boost in the z-direction.  This procedure can 
be repeated for boosts in the x- and y-directions with similar results.  

After performing boosts and rotations in the condition given by Eq. 
(\ref{eq33}), the only remaining free parameters for $B_c$ from Eq. 
(\ref{eq37a}) are $a$ and $c_\mu$.  No other restriction is available 
to eliminate these free parameters so they will appear in the first 
order equation.  These free parameters are allowed because of the 
added flexibility in the state function that is afforded by the 
freedom to select any phase function $\phi$ to offset non-invariant 
terms generated by the transformation of the constant matrix $B_c$.  
The free parameters $c_\mu$ and $a$ will appear in the first order 
equation making it more general than the Dirac equation given by Eq. 
(\ref{eq30d}).  Thus we have a generalized Dirac equation
\begin{equation}
\left[ \gamma^\mu \partial_\mu + aI + c_\mu \gamma^\mu \right]\psi(x,t)
=0\ ,
\label{eq50}
\end{equation}

\noindent
which satisfies the principles of relativity and analyticity.  We call this 
new fundamental Poincar\'e invariant equation the generalized Dirac equation 
because the original Dirac equation (see Eq. \ref{eq30d}) is obtained from it 
as a special case when the phase function $\phi$ is set to zero.  To be more 
specific, the standard results are obtained when the phase function $\phi=0$ 
requires $\zeta_\mu=0$ which in turn requires the free parameters to vanish 
$c_\mu=0$.   

Other generalized Dirac equations were published in the literature [24-28],
however, they are completely different than Eq. (\ref{eq50}).  Kruglov [27,28] 
generalized a Dirac equation to account for a particle with two mass states, 
and the resulting 'generalized Dirac equation' was a second order differential 
equation.  As such the equation resembles rather the Klein-Gordon equation than 
the Dirac equation.  Actually, the equation has the additional first order 
parameter found in the Dirac equation and, therefore, it can be considered as 
a sum of Dirac and Klein-Gordon equations.  The 'generalized Dirac equation' 
described by Nozari [29] follows from a generalized uncertainty principle and 
exists in the context of spacetime with an assumed minimal distance on the order 
of the Planck length.  This equation is not Poincar\'e invariant and thus its 
meaning is different than Eq. (\ref{eq50}).  More discussion relevant to the 
comparison of our results to those obtained previously will be given in section 7.  
%

%%%%%%%%%%%%%%%%%%%%%%%%%%%%%%%%%%%%%%%%%%%%%%%%%%%%%%%%%%%%%%%%%%%%%%%%%%%
\section{Generalized Klein-Gordon equation}

Having derived the generalized Dirac equation, we may now use it to obtain the 
generalized Klein-Gordon equation.  However, before this formally done, first 
we write the generalized Dirac equation in its alternate form.  We multiply 
Eq. (\ref{eq50}) by $i\gamma_0$, use the momentum operator $\hat{p}_i = - i
\partial_i$, energy operator $\hat{\varepsilon}=i\partial_t$, and separate 
the time derivative from the space derivative.  The result is
\begin{equation}
\hat{\varepsilon}\psi =\mathcal{H}\psi= \left[\alpha_i \hat{p_i}+
\tilde{a}\beta+\tilde{c_0}I+\alpha_i\tilde{c_i}\right]\psi\ ,
\label{eq51}
\end{equation}

\noindent
where $\alpha_i=\gamma_0\gamma_i$, $\beta=\gamma_0$, $\tilde{a}=-ia$, $\tilde{c_0}
=-ic_0$, and $\tilde{c_i}=-ic_i$.  In this form the RHS can be identified as the 
Hamiltonian $\mathcal{H}$ acting on the wavefunction $\psi$.  Since the eigenvalues 
of $\mathcal{H}$ must be real and thus $\mathcal{H}$ must be Hermitian.  Thus, 
$\alpha_i$ and $\beta$ are Hermitian and the Hermiticity of the Hamiltonian is 
satisfied if $\tilde{a}$ and $\tilde{c_\mu}$ are real.

Now, a second order equation such as the Klein-Gordon equation [4,5] can be 
constructed by applying the operators of the generalized Dirac equation to 
the wavefunction a second time.  Starting with Eq. (\ref{eq51}) and making 
substitutions $\tilde{a}=m_0$, $\tilde{c_0}=-\tilde{\varepsilon}$, 
$\tilde{c_j}=\tilde{p}_j$, and $\hat{p}_j=-i\partial_j$, we obtain the 
following first order equation
\begin{equation}
i \partial_t \psi = \left[ -i \alpha_j \partial_j + m_0 \beta - 
\tilde{\varepsilon} + \alpha_j \tilde{p}_j \right] \psi\ .
\label{eq52}
\end{equation}

\noindent
After applying the operators to the wavefunction a second time, the 
equation becomes 
\[
-\partial_t^2 \psi= -\partial_j^2 \psi + m_0^2 \psi + \tilde{p}_j^2 
\psi + \tilde{\varepsilon}^2 \psi-2i \tilde{p}_j \partial_j \psi - 2m_0
\tilde{\varepsilon} \beta \psi
\]
\begin{equation}
 +2i \tilde{\varepsilon} \alpha_j \partial_j \psi - 2\tilde{\varepsilon} 
 \tilde{p}_j \alpha_j \psi\ , 
\label{eq53}
\end{equation}

\noindent
which is the generalized Klein-Gordon equation.  The terms are all Hermitian 
and the equation is Poincar\'e invariant.  

%%%%%%%%%%%%%%%%%%%%%%%%%%%%%%%%%%%%%%%%%%%%%%%%%%%%%%%%%%%%%%%%%%%%%%%%%
\section{Non-relativistic limits of generalized Dirac equation}

\subsection{Generalized Pauli-Schr\"odinger equation}

In order to consider the non-relativistic limit, we replace the natural units 
in the generalized Dirac equation by the physical units.  The resulting equation
shows explicitly the speed of light $c$, which is required to take the limit.
The generalized Dirac equation in its physical units can be written as
\begin{equation}
i\partial_t \psi =\mathcal{H}= \left[c \alpha_i \hat{p_i} + 
c\tilde{a}\beta+c\tilde{c_0}I+c\alpha_i\tilde{c_i}\right]\psi\ .
\label{eq54}
\end{equation}

Introducing the electromagnetic four potential $A^\mu=\left(A_0(x), A_j(x)\right)$,
which can be incorporated into the generalized Dirac equation via the minimal 
coupling
\begin{equation}
p^\mu\rightarrow p^\mu-{e\over c} A^\mu\equiv\Pi^\mu\ ,
\label{eq55}
\end{equation}

\noindent
where $\Pi^\mu$ is the kinetic momentum and $e$ is the electron charge.  Then, 
the generalized Dirac equation with the electromagnetic potential is
\begin{equation}
i\partial_t \psi = \left[c \alpha_i \left(\hat{p_i}-{e\over c}A_i\right) + 
eA_0+c\tilde{a}\beta+c\tilde{c_0}I+c\alpha_i\tilde{c_i}\right]\psi\ .
\label{eq56}
\end{equation}

The four-component spinor $\psi$ can be decomposed into a pair of two-component 
spinors $\tilde{\phi}$ and $\tilde{\chi}$
\begin{equation}
\psi=\left(\begin{array}{c}
\tilde{\phi} \\
\tilde{\chi}
\end{array}
\right)\ .
\label{eq57}
\end{equation}

\noindent
Inserting explicit representations for the matrices $\alpha_i$, $\beta$, and 
$I$, we write the generalized Dirac equation as
\[
i\partial_t
\left(\begin{array}{c}
\tilde{\phi} \\
\tilde{\chi}
\end{array}
\right)
 = c \left(\begin{array}{c}
\sigma_i \Pi_i \tilde{\chi} \\
\sigma_i \Pi_i \tilde{\phi}
\end{array}
\right)+
eA_0
\left(\begin{array}{c}
\tilde{\phi} \\
\tilde{\chi}
\end{array}
\right)
+
c\tilde{a}
\left(\begin{array}{c}
\tilde{\phi} \\
-\tilde{\chi}
\end{array}
\right)
\]
\begin{equation}
+c\tilde{c_0}
\left(\begin{array}{c}
\tilde{\phi} \\
\tilde{\chi}
\end{array}
\right)
+c\tilde{c_i}
\left(\begin{array}{c}
\sigma_i\tilde{\chi} \\
\sigma_i\tilde{\phi}
\end{array}
\right)\ .
\label{eq58}
\end{equation}

The non-relativistic limit of the generalized Dirac equation requires making 
some assumptions about the value of the free parameters $\tilde{a}$ and 
$\tilde{c_\mu}$.  In the standard Dirac equation $\tilde{a}=m_0c$, where 
$m$ is the mass of the particle and $c$ is the speed of light.  We shall assume 
this value for $\tilde{a}$ in order to interpret the third term on the RHS as 
a rest mass energy.  The similarity of the fifth term to the first term is 
suggestive of a momentum interpretation while the similarity between the 
fourth term and left hand side term suggests an energy interpretation.  
We shall thus set $c\tilde{c_0}=-\tilde{\varepsilon}$, where the negative 
sign is chosen to match the sign of $\varepsilon$ as it will appear on the 
left.  Now if we assume $m_0c^2$ is the largest energy, the spinor state 
function may be further split into two parts
\begin{equation}
\left(\begin{array}{c}
\tilde{\phi} \\
\tilde{\chi}
\end{array}
\right)
=
\left(\begin{array}{c}
{\phi} \\
{\chi}
\end{array}
\right)\exp\left[-im_0c^2t\right]\ .
\label{eq59}
\end{equation}

\noindent
The generalized Dirac equation is now
\[
i\partial_t
\left(\begin{array}{c}
{\phi} \\
{\chi}
\end{array}
\right)
 = c \left(\begin{array}{c}
\sigma_i \Pi_i {\chi} \\
\sigma_i \Pi_i {\phi}
\end{array}
\right)+
eA_0
\left(\begin{array}{c}
{\phi} \\
{\chi}
\end{array}
\right)
-2m_0c^2
\left(\begin{array}{c}
0 \\
{\chi}
\end{array}
\right)
\]
\begin{equation}
-\tilde{\varepsilon}
\left(\begin{array}{c}
{\phi} \\
{\chi}
\end{array}
\right)
+c\tilde{c_i}
\left(\begin{array}{c}
\sigma_i{\chi} \\
\sigma_i{\phi}
\end{array}
\right)\ .
\label{eq60}
\end{equation}

If we consider the second component of the equation when kinetic and 
potential energies are small compared to the rest mass energy, i.e., 
when $\left|i\partial\chi / \partial t \right|\ll \left| m_0c^2
\chi \right|$, $\left|\tilde{\varepsilon \chi} \right| \ll \left| 
m_0c^2\chi \right|$, and $\left|eA_0 \chi \right| \ll 
\left|m_0 c^2 \chi \right|$ then we find that
\begin{equation}
\chi={\sigma_j \left(\Pi_j + \tilde{c_j}\right) \over 2m_0c }\phi\ .
\label{eq61}
\end{equation}

\noindent
This relation can be substituted into the first component Eq. (\ref{eq60})
and, as a result, the generalized Pauli-Schr\"odinger equation is obtained
\begin{equation}
i \partial_t \phi = \left[ {1 \over 2m_0} \left( p_j - {e \over c} 
A_j + \tilde{c_j}\right)^2 - {{e\hbar \sigma_j} \over {2m_0c}} B_j + 
{{\sigma_j \epsilon_{jkl}} \over {2m_0}} \partial_k \tilde{c_l} + 
e A_0 - \tilde{\varepsilon} \right] \phi\ .
\label{eq62}
\end{equation}

\noindent
Note that for a constant $\tilde{c_j}$ the curl term vanishes leaving one difference 
from the original Pauli-Schr\"odinger equation  [17].  Here the $\tilde{c_j}$ 
term contributes to the time rate of change of $\phi$ in the same way as the momentum 
does.

\subsection{Generalized L\'evi-Leblond equation}

Returning to the first component equation of (\ref{eq60}), we use the 
energy eigen-operator $\hat{\varepsilon}=i\partial_t$ and the eigen-equation 
$\hat{\varepsilon}\psi=\varepsilon \psi$, and set the electromagnetic potential 
and charge to zero, we obtain the first half of the generalized L\'evy-Leblond 
equation.  Doing the same with equation (\ref{eq61}), we get the second half 
of the generalized L\'evy-Leblond equation.  Finally, we can write the 
generalized L\'evy-Leblond equation [22] in its full form as 
\begin{equation}
(\varepsilon +\tilde{\varepsilon}) \phi = \sigma_j \left( p_j + \tilde{c_j} 
\right) \chi\ ,
\label{eq63a}
\end{equation}
and
\begin{equation}
2m_0 \chi = \sigma_j \left( p_j + \tilde{c_j} \right) \phi\ .
\label{eq63b}
\end{equation}

\noindent
Derivation of this generalized L\'evy-Leblond equation was already done by 
Huegele et al. [22], who used the principle of relativity and the principle 
of analyticity, as well as the extended Galilei group.  They showed that the
equation is Galilean invariant and discussed its possible applications.
%    

%%%%%%%%%%%%%%%%%%%%%%%%%%%%%%%%%%%%%%%%%%%%%%%%%%%%%%%
\section{Generalized Dirac equation and families of elementary particles}

We now examine motions of a free particle described by the generalized Dirac 
equation (see Eq. \ref{eq60}).  Let us look for stationary states $\psi(x)$ 
that have no time dependence, so the wavefunction is separable from the 
stationary state such that
\begin{equation}
\psi(x,t)=\psi(x)exp[-i\varepsilon t]\ .
\label{eq64}
\end{equation}

\noindent
Under this condition the generalized Dirac equation is a time independent 
equation of the stationary state
\begin{equation}
\varepsilon \psi(x) =  \left[ -i \alpha_j \partial_j + m_0 \beta - 
\tilde{\varepsilon}+ \alpha_j \tilde{p}_j \right] \psi(x)\ .
\label{eq65}
\end{equation}

\noindent
Splitting the four-component spinor into a pair of two-component spinors and 
using explicit representations for the matrices $\alpha$ and $\beta$, we write 
Eq. (\ref{eq65}) as the following pair of equations
\begin{equation}
\varepsilon \phi = \sigma_j p_j \chi + m_0 \phi - \tilde{\varepsilon} 
\phi + \sigma_j \tilde{p}_j \chi\ ,
\label{eq66a}
\end{equation}
and
\begin{equation}
\varepsilon \chi = \sigma_j p_j \phi - m_0 \chi -\tilde{\varepsilon}
\chi + \sigma_j \tilde{p}_j \phi\ .
\label{eq66b}
\end{equation}

States with a defined momentum $p_j$ 
\begin{equation}
\left( \begin{array} {c} \phi \\ \chi \end{array} \right)
= \left( \begin{array} {c} \phi_0 \\ \chi_0 \end{array} \right) 
exp\left[ i\hat{p_j} x_j \right]\ ,
\label{eq67}
\end{equation}

\noindent
can be substituted into Eqs (\ref{eq66a})-(\ref{eq66b}).  Moreover, we replace 
the operator $\hat{p_j}$ with its eigenvalues $p_j$, and obtain
\begin{equation}
\left( \varepsilon + \tilde{\varepsilon} - m_0 \right) \phi_0 - 
\sigma_j \left( p_j + \tilde{p}_j \right) \chi_0 = 0\ ,
\label{eq68a}
\end{equation}
and
\begin{equation}
-\sigma_j \left( p_j + \tilde{p}_j \right) \phi_0+\left( \varepsilon
+\tilde{\varepsilon} + m_0 \right) \chi_0 = 0\ .
\label{eq68b}
\end{equation}

\noindent
This system of equations admits non-trivial solution only when
\begin{equation}
\left( \varepsilon +\tilde{\varepsilon} - m_0 \right) \left(\varepsilon 
+\tilde{\varepsilon}+ m_0 \right) - \left[ \sigma_j \left( p_j + 
\tilde{p}_j \right) \right]^2=0\ ,
\label{eq69a}
\end{equation}

\noindent
which can be simplified to
\begin{equation}
(\varepsilon+\tilde{\varepsilon})^2 - m_0^2 - \left( p_j + \tilde{p}_j 
\right)^2=0\ .
\label{eq69b}
\end{equation}

Comparison of this energy-momentum relationship to that of Special Theory of
Relativity (STR) shows the presence of two additional terms $\tilde{\varepsilon}$ 
and $\tilde{p}_j$ in the former.  If $\tilde{\varepsilon} = 0$ and $\tilde{p}_j 
= 0$, then the standard STR energy-momentum relationship, which also underlines
the standard Dirac equation (see Eq. \ref{eq50}), is recovered.  To reconcile 
both relationships, we introduce $E = \varepsilon + \tilde{\varepsilon}$ and 
$P_j = p_j + \tilde{p}_j$, and recognize the fact that both relationships 
describe the same elementary particle with spin 1/2 but with different masses.
With $m > m_0$ being a mass of the more massive particle, we write the resulting 
energy-momentum relationship for the generalized Dirac equation as 
\begin{equation}
E^2 - m^2 - P_j^2 = 0\ ,
\label{eq70}
\end{equation}

\noindent
which, as expected, is consistent with STR.

The main obtained result is that the standard and generalized Dirac equations 
describe the same elementary particle with spin 1/2 but with different masses; 
the mass $m$ in the generalized Dirac equation is always larger than the mass 
$m_0$ in the standard Dirac equation.  Since the standard Dirac equation is 
derived with the phase function $\phi = 0$ and the generalized Dirac equation
corresponds to $\phi = 0$, we may draw an important conclusion, namely, that 
{\it non-zero phase functions automatically account for the same particles but 
with different masses.}  A significant physical consequence is that the existence 
of the three currently known families of elementary particles in the Standard 
Model of particle physics [22] may be theoretically accounted for by choosing 
different phase factors.  To the best of our knowledge this is a new idea, which 
requires comparison to the previously suggested solutions of this problem.

\section{Comparison to the previous work}

The existence of three families of elementary particles in the Standard 
Model [22] confirmed by high energy physics experiments is one of the 
main longstanding problems in modern physics. A number of interesting 
ideas were proposed to explain the origin of the three families [23-31] 
but to the best of our knowledge the problem remains still unsolved.  

Barut and his coworkers [24,25] obtained a second-order dynamical equation 
describing elementary particles with two mass states and suggested that this 
approach gives a unified description of electrons and muons.  Similar equation
was considered and analyzed by Kruglov [26].  However, in his more recent 
work Kruglov extended Barut's work by obtaining a generalized Dirac equation 
for a 20-component [27] and 16-component [28] wavefunctions and showed that 
such functions can represent fermions with two mass states; it must be pointed 
out that Kruglov's work described in [27] is based on the assumption that 
Lorentz invariance is violated.  

In theoretical models with extra dimensions, the Dirac equation is also 
considered and its three zero modes introduced by some background fields 
of nontrivial topology are typically identified with the three generations
of elementary particles [30,31].  A new mechanism for the origin of the 
three fermion generations was recently proposed by Kaplan and Sun [32],
who considered fifth dimensional space-time as a topological insulator 
and suggested that the three generations of leptons and quarks correspond 
to surface modes in their theory.  These are interesting theoretical 
suggestions, however, their main problem is that so far there is no
experimental evidence for the existence of any extra dimensions.

Our novel results presented in this paper suggest that there is 
another way to account for the three families of elementary particles
in the Standard Model [23,33], namely, by considering non-zero phase 
functions in the relativistic models.  If this is indeed the case, 
then it would be possible, at least in principle, to determine 
theoretically a number of families of elementary particles by 
finding how many physically meaningful phase functions are allowed.  
However, such investigation is out of the scope of the present paper.

\section{Conclusion}

We developed a method that allowed us to search for new Poincar\'e 
invariant dynamical equations describing free elementary particles,
which are represented by four component spinor wavefunctions.  The
original Dirac equation was obtained and shown to be the only Poincar\'e 
invariant dynamical equation for 4-component spinor wavefunctions 
when the phase function is $\phi=0$.  We also considered non-zero
phase functions and demonstrated that they give new Poincar\'e 
invariant generalized Dirac and Klein-Gordon equations, which reduce
to the original Dirac and Klein-Gordon equations when the phase function 
$\phi$ is set to zero.  These are important results as they show that 
other fundamental (Poincar\'e invariant) equations do exist in Minkowski 
space-time.

To validate our generalized Dirac equation, we derived the previously 
obtained generalized L\'evy-Leblond equation and a new generalized 
Pauli-Schr\"odinger equation by taking the non-relativistic limits 
of the generalized Dirac equation.  We also demonstrated that both 
generalized equations could be reduced to the standard L\'evy-Leblond
and Pauli-Schr\"odinger equations by taking the phase function to be 
zero.  

Our results clearly show that the main difference between the original
and generalized Dirac equations is that they are obtained with zero and
non-zero phase functions, respectively, and that these equations describe 
free elementary particle with spin 1/2, which have all other physical 
properties the same except their masses.  The fact that the generalized 
Dirac equation describes elementary particles with larger masses is used 
to suggest that non-zero phase functions may account for the existence 
of three families of elementary particles in the Standard Model.  This 
suggestion significantly differs from those previously made to account 
for the three families of particle physics. 

\bigskip\noindent
{\bf ACKNOWLEDGEMENTS.}  
Z.E.M. acknowledges the support of this work by the Alexander von 
Humboldt Foundation and by The University of Texas at Arlington through 
its Faculty Development Program.

\end{document}